\documentclass[preprint,5p,times,twocolumn]{elsarticle}
\usepackage[T1]{fontenc}
\usepackage[utf8]{inputenc}

\usepackage{amsmath}
\usepackage{amssymb}

\usepackage{color}
\usepackage{soul}

\newcommand{\bm}[1]{\mbox{\boldmath $ #1 $ \unboldmath} \!\!}

\newcommand{\Karman}{\textsc{K\'arm\'an} }
\newcommand{\von}{\textsc{von} }
\newcommand{\Von}{\textsc{Von} }
\newcommand{\Liepmann}{\textsc{Liepmann} }
\newcommand{\Modified}{\textsc{Modified} }
\newcommand{\modified}{\textsc{modified} }

\newcommand{\Gaussian}{\textsc{Gaussian} }

\usepackage{mathrsfs}
\usepackage{amsbsy}

\newcommand{\abl}{\text{d}}

\newcommand{\N}{\mathscr{N}}  
\renewcommand{\L}{\mathcal{L}} 
\newcommand{\G}{\mathcal{G}} 

\newcommand{\bpm}{\begin{pmatrix}}
\newcommand{\epm}{\end{pmatrix}}
\usepackage{mathrsfs}
\usepackage{subfig}
\usepackage{pdflscape}

\journal{Computers \& Fluids}

\begin{document}

\begin{frontmatter}



\title{Analytical reconstruction of \\ isotropic turbulence spectra \\
		based on the Gaussian transform}


\author[atta]{A. Wohlbrandt\footnote{Corresponding Author: attila.wohlbrandt@dlr.de,\\ Telephone +49 30 310006-21, Fax +49 30 310006-39}}
\author[asta]{N. Hu}
\author[atta]{S. Gu\'erin}
\author[asta]{R. Ewert}
\address[atta]{Institute of Propulsion Technology, Engine Acoustics Department\\ German Aerospace Center (DLR), M\"uller-Breslau-Str.8, 10623 Berlin, Germany}
\address[asta]{Institute of Aerodynamics and Flow Technology, Technical Acoustics\\ German Aerospace Center (DLR), Lilienthalplatz 7, 38108 Braunschweig, Germany}

\begin{abstract}
The Random Particle Mesh (RPM) method used to simulate turbulence-induced broadband noise in several aeroacoustic applications is extended to realise isotropic turbulence spectra.
With this method turbulent fluctuations are synthesised by filtering white noise with a \Gaussian filter kernel that in turn gives a \Gaussian spectrum. The \Gaussian function is smooth and its derivatives and integrals are again \Gaussian functions. The \Gaussian filter is efficient and finds wide-spread applications in stochastic signal processing. 
However in many applications \Gaussian spectra do not correspond to real turbulence spectra. Thus in turbo-machines the \textsc{von K\'arm\'an, Liepmann}, and \modified \von \Karman spectra are more realistic model spectra.
In this note we analytically derive weighting functions to realise arbitrary isotropic solenoidal spectra using a superposition of weighted \Gaussian spectra of different length scales. The analytic weighting functions for the \von \Karman, the \Liepmann, and the \modified \von \Karman spectra are derived subsequently. Finally a method is proposed to discretise the problem using a limited number of \Gaussian spectra. The effectivity of this approach is demonstrated by realising a \von \Karman velocity spectrum using the RPM method. 

\end{abstract}

\end{frontmatter}
\section{Introduction}
A stochastic noise signal of a certain spectral shape can be generated by convolution of a white noise signal by a filter kernel of an appropriate shape~\cite{Ewert_CAA_2011}.

One of the most common filter kernels is the \Gaussian filter kernel that realises a \Gaussian spectrum. The \Gaussian filter is very simple and time efficient as it has benefitial characteristics: its derivatives and integrals are again of \Gaussian shape; the filtering decouples in multi-dimensional space and fast filter methods are available, such as \textsc{Purser}~\cite{purser_numerical_2003} and \textsc{Young \& Van-Vliet}~\cite{young_recursive_1995} filters.

But \Gaussian spectra seldom represent the physics of turbulence. Here more elaborate spectra are needed, such as \textsc{Kolmogorov}, \von \Karman or \Liepmann spectra. For these spectra the filter kernels are very complicated  to use and they are fully coupled in space, as shown by Dieste and Gabard~\cite{Dieste_Random_2012}.


Siefert et al.~\cite{Siefert_Sweeping_2009}  succeeded in realising the \textsc{Kolmogorov} spectrum using the superposition of Gaussian spectra weighted manually. Others have adopted this method for other kinds of spectra, e.g. just recently Gea-Aguilera et al.~\cite{Gea-Aguilera_Synthetic_2015} and Kim et al.~\cite{Kim_Advanced_2015} published their findings. Note that the here presented method has already been presented by the authors on conferences, but not derived in detail~\cite{Envia_AA-39_2014,Rautmann_Generic_2014}.

The objective of this note is to provide a theoretical background for determining the appropriate analytical weighting function by means of \Gaussian transform~\cite{Alecu_Gaussian_2005}. The analytical weighting function is derived for the \von \textsc{K\'arm\'an}, the \Liepmann and the \modified \von \Karman spectra. 
Furthermore, an efficient method is proposed to discretise the weighting function with a limited number of Gaussian spectra. Suggestions are made to choose the number of filters and their length scales. 
As illustration, the realised velocity spectrum using the Random Particle Mesh (RPM) method~\cite{Ewert_CAA_2011} is compared to the analytically derived velocity spectrum.  

\section{Method - \Gaussian Transformation}
\subsection{Turbulence spectra} 
The most popular models for isotropic turbulence are the \textsc{von K\'arm\'an, Liepmann,} and \textsc{modified von K\'arm\'an} models~\cite{Atassi_Effect_2008}.

\subsubsection{The \von \Karman Spectrum} 
The \von \Karman spectrum is commonly used to represent homogeneous isotropic turbulence. It satisfies the energy law distribution of $k^4$
for the large eddies which contain most of the energy and reproduces the $-5/3$ - law in the inertial subrange. The energy spectrum is given by
	\begin{equation}
		E_K (\hat k) = \frac{55 }{9\pi}u_t^2\Lambda\frac{\hat k^4}{(1+\hat k^2)^{17/6}}, \label{eq:KarmanEnergy}
	\end{equation}
with the mean turbulent velocity $u_t$ related to the turbulent intensity $T_u$ and the mean flow velocity $u_0$ by $u_t^2 = (T_u \cdot u_0)^2$, the integral length scale $\Lambda$, and the reduced wavenumber $\hat k$ defined by $\hat k = k^* /k_e$, where $k^* = k \Lambda$, and $k_e = \frac{\sqrt{\pi}\Gamma(5/6)}{\Gamma(1/3)}$. 
\subsubsection{The \Liepmann Spectrum} 
\Liepmann determined turbulence longitudinal correlation coefficients from measurements and found that they can be approximated by an exponential law $f(x) = \exp\left(\frac{-x}{\Lambda}\right)$. The resulting model spectrum is given as \cite{Atassi_Effect_2008,Hinze_Turbulence_1975}:
\begin{align}
E_L(k^*) = \frac{8 u_t^2 \Lambda}{\pi} \frac{{k^*}^4}{(1+{k^*}^2)^3}. \label{eq:LiepmannSpectrum}
\end{align}
The \Liepmann spectrum is comparable in shape to the \von \Karman spectrum. 

\subsubsection{The \modified \von \Karman spectrum}
According to Bechara~\cite{Bechara_Stochastic_1994} the \von \Karman spectrum can be modified to be representative over the entire wavenumber range including the dissipation subrange:

\begin{align}\label{eq:modifiedKarmanEnergy}
		E_{M} (\hat k) = E_K(\hat k) \exp\left(-2\frac{k^2}{k_d^2}\right) 
\end{align}
with the \textsc{Kolmogorov} wavenumber $k_d = \left(\frac{\epsilon}{\nu^3}\right)^{1/4}$, where $\epsilon$ is the specific dissipation rate and $\nu$ is the eddy viscosity.

\subsection{Weighting function}
According to Ewert et al.~\cite{Ewert_CAA_2011} filtering of a white noise field
with a \Gaussian filter kernel of a specific length scale realises a \Gaussian
spectrum of the form
\begin{equation}
	E_G (k) = \frac{4 u_t^2 \Lambda}{\pi^3} {k^*}^4 e^{\frac{-{k^*}^2}{\pi}}. \label{eq:GaussEnergy}
\end{equation}

For convenience we introduce a new spectrum $e_G$ such that its integral over the wavenumber range is one, i.e.
\begin{gather}
\int\limits_0^\infty e_G(k) \abl k= 1 \qquad
\Rightarrow E_G(k) = \frac{3}{2}u_t^2 e_G(k).\label{eq:norm}
\end{gather}

We are looking for a weighting function $f(l,\Lambda)$ to realise an arbitrary spectrum $e(k)$ of integral length scale $\Lambda$ by means of a superposition of \Gaussian spectra $e_G$ of length scales $l$:
\begin{align}
e(k,\Lambda) &= \int\limits_0^\infty f(l,\Lambda) e_G(k,l) \abl l. \notag \\
\intertext{Using Eq.~\eqref{eq:GaussEnergy} and \eqref{eq:norm} yields the following solution:}
\label{eq:weighting}
e(k,\Lambda) &= \int\limits_0^\infty f(l,\Lambda) \frac{8}{3\pi^3} l^5 k^4 \exp\left(-\frac{k^2l^2}{\pi}\right)\abl l.
\end{align}
Note that only the weighting function $f(l,\Lambda)$ depends on the integral length scale $\Lambda$. In the following we drop $\Lambda$ in the expression of the weighting function $f$ and write $f(l,\Lambda) = f(l)$.

A parameter $\sigma$ is introduced to write Equation~\eqref{eq:weighting} in a suitable manner for \Gaussian transform as defined by Alecu et al.~\cite{Alecu_Gaussian_2005}. This parameter verifies the two following relationships:
\begin{align}
&&l^2 = \frac{\pi}{2 \sigma^2} && \text{and} && \frac{\abl l}{\abl \sigma^2} = - \frac{\sqrt{\pi}}{2\sqrt{2} \sigma^3}. && \notag
\end{align}
Equation~\eqref{eq:weighting} rewrites:
\begin{align}
\underbrace{\frac{e(k)}{k^4}}_{p(k)} &=  \int\limits_0^\infty 
			\underbrace{f\left(\sqrt{\frac{\pi}{2}}\frac{1}{ \sigma}\right) \frac{\sqrt{\pi}}{3\sqrt{2}\sigma^7}}_{G(\sigma^2)}
			\underbrace{\frac{1}{\sqrt{2\pi\sigma^2}}  \exp\left(-\frac{k^2}{2 \sigma^2}\right)}_{\N(k|\sigma^2)}\abl \sigma^2. \label{eq:gaussTransform}
\end{align}
According to Alecu et al., $p(k)$ is a zero-mean generic symmetric distribution, $G(\sigma^2)$ is the mixture function and $\N(k|\sigma^2)$ is the zero-mean Gaussian distribution. They define the \Gaussian transform $\G$ as the operator which transforms $p(x)$ into $G(\sigma^2)$. The inverse \Gaussian transform $\G^{-1}(G(\sigma^2)) = p(x)$ is simply given by Equation~\eqref{eq:gaussTransform}.

From Eq.~\eqref{eq:gaussTransform} the weighting function $f(l)$ is given as
\begin{align}
	f\left(l = \sqrt{\frac{\pi}{2}}\frac{1}{\sigma}\right) = \frac{3\sqrt{2}\sigma^7}{\sqrt{\pi}} G(\sigma^2).
\end{align}

\subsubsection{\Von \Karman weighting function}
With the \von \Karman spectrum given in Eq.~\eqref{eq:KarmanEnergy} the left-hand side of Equation~\eqref{eq:gaussTransform} becomes
\begin{align}\label{eq:KarmanEnergy2}
p(k) 
	= \frac{110}{27\pi}\Lambda^5 k_e^{5/3}\frac{1}{(k_e^2+k^2\Lambda^2)^{17/6}}.
\end{align}
The direct \Gaussian Transform is given by Alecu et al.~\mbox{\cite[Eq.(4)]{Alecu_Gaussian_2005}}:
\begin{align}\label{eq:dirGaussTrans}
\G\left(p\left(\sigma\right)\right) = \frac{1}{\sigma^2} \sqrt{\frac{\pi}{2\sigma^2}}\Bigg(\L^{-1}\Big(p(\sqrt{s})\Big)(t)\Bigg)_{t=\frac{1}{2\sigma^2}}.
\end{align}
where $ \L^{-1}$ is the inverse \textsc{Laplace} transform. Using the relation
\begin{align}
\L^{-1} \left(\frac{1}{(p-\alpha)^{n}}\right)(t) = \frac{e^{\alpha t} t^{n-1}}{\Gamma(n)},\label{eq:KarmanInvLaplace}
\end{align}
where $\Gamma(n) = (n-1)!$ is the gamma function, we find for Eq.~\eqref{eq:KarmanEnergy2}
\begin{align}
\G(p(\sigma))
&= \frac{55 }{54 \sqrt{\pi}  \Gamma(17/6)} \sqrt[3]{\frac{k_e^5}{2 \sigma^{20} \Lambda^2}}
\exp\left(-\frac{k_e^2 }{2 \Lambda^2 \sigma^2}\right)\label{eq:KarmanGauss}
\end{align}
and the weighting function for the \von \Karman spectrum is given by
\begin{align}\label{eq:weightKarman}
\boxed{f_K(l)= \frac{55 }{18 \Gamma(17/6)\sqrt{\pi}} \sqrt[3]{\frac{k_e^5}{\pi \Lambda^2 l}}
\exp\left(-\frac{k_e^2 l^2}{\pi\Lambda^2}\right)}
\end{align}

\begin{figure}[h]
	\centering
	\includegraphics[width=0.5\textwidth]{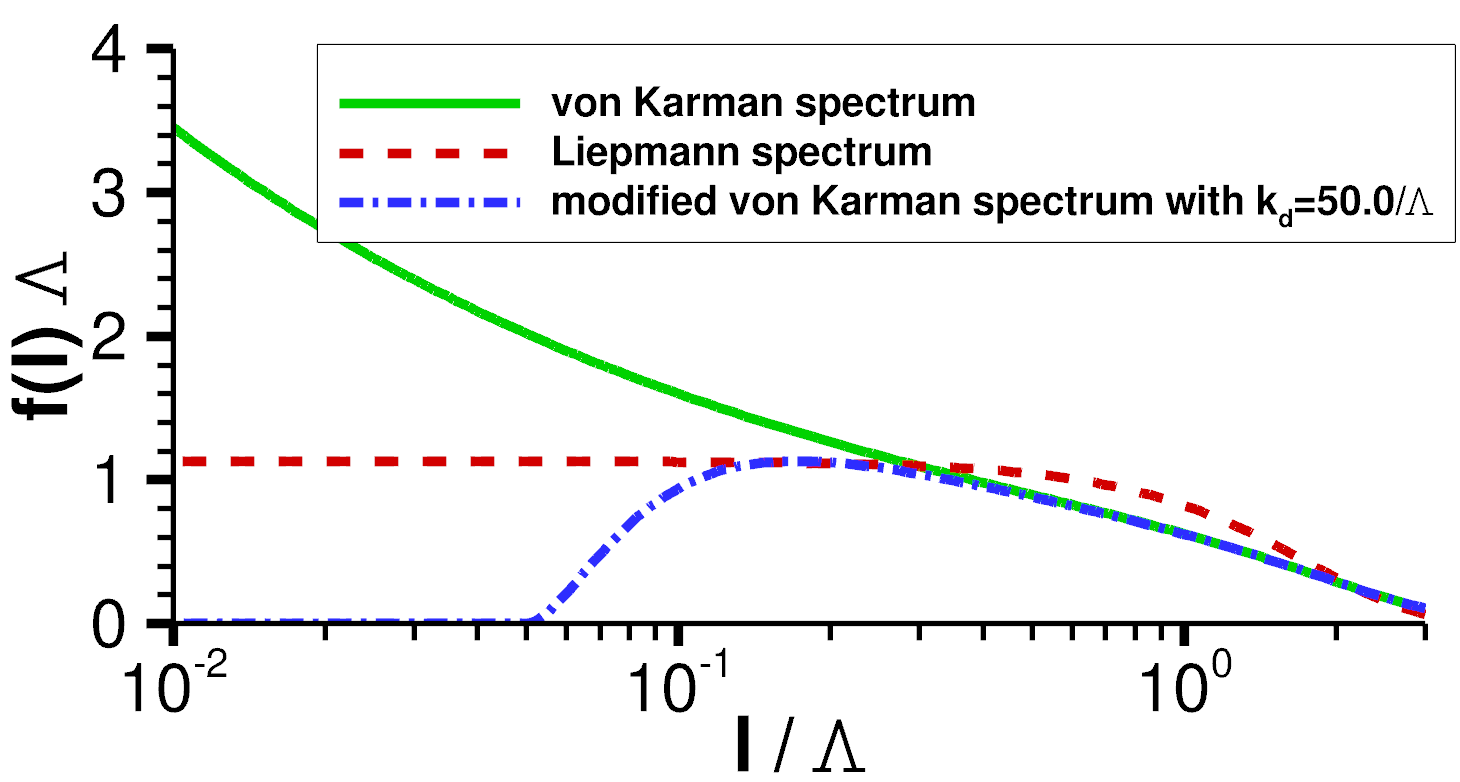}
	\caption{The weighting function $f(l)$ is used to realise typical turbulence spectra of length scale $\Lambda$ using a superposition of \Gaussian spectra of various length scales $l$. The analytical weighting functions for the \von \Karman (\textcolor[rgb]{0,0.8,0}{solid}), the \Liepmann  (\textcolor{red}{dashed}), and the \modified \von \Karman (\textcolor{blue}{dash-dot}) spectra are shown.\label{fig:weight}}
\end{figure}

\subsubsection{\Liepmann weighting function}


With the \Liepmann spectrum given in Eq.~\eqref{eq:LiepmannSpectrum} the left-hand side of Equation~\eqref{eq:gaussTransform} becomes
\begin{align}
p(k) 
	= \frac{16 \Lambda^4}{3 \pi } \frac{\left(\frac{1}{\Lambda^2}\right)^{5/2}}{(\frac{1}{\Lambda^2}+k^2 )^3}
\end{align}
This is of the form of the generalised Cauchy distribution shown in appendix of Ref.~\cite{Alecu_Gaussian_2005} with $\nu = 2.5$ and $b = \frac{1}{\Lambda}$. We identify
\begin{align}
p(x) = \frac{16 \Lambda^4}{3 \pi } \frac{\sqrt{\pi} \Gamma(2.5)}{2} p_x^C ( x|\nu,b)
\end{align}
The \Gaussian transform of $p_x^C ( x|\nu,b)$ is given in \cite{Alecu_Gaussian_2005} as 
\begin{align}
G_{X|\nu,b}(\sigma^2) &= \frac{b^5 2}{\sigma^2 \sqrt{2\sigma^2}\Gamma(2.5)} \left( \mathcal{F}^{-1}\left(\left(b^2+i\omega\right)^{-3}\right)\right)_{t=\frac{1}{2\sigma^2}}
\end{align}
So the \Gaussian transform of $p(x)$ is
\begin{align}
G(\sigma^2) 
&= \frac{ 2}{3\Lambda \sigma^6 \sqrt{2\pi \sigma^2}}  e^{-\frac{1}{2\Lambda^2 \sigma^2}}
\end{align}
Finally, this yields the weighting function for a \Liepmann spectrum as
\begin{align}\label{eq:weightLiepmann}
\boxed{f_L(l)= \frac{ 2}{\pi \Lambda}  e^{-\frac{l^2}{\pi \Lambda^2 }}}
\end{align}
This function is plotted in Fig.~\ref{fig:weight}.

\subsubsection{\Modified \von \Karman weighting function}

With the \modified \von \Karman spectrum given in Eq.~\eqref{eq:modifiedKarmanEnergy} the left-hand side of Equation~ \eqref{eq:gaussTransform} becomes
\begin{align}
p(k) &= \frac{1}{k^4}\frac{2 E_{M}(k)}{3 u_t^2} \notag\\ 
	&= \frac{110}{27\pi}\Lambda^5 k_e^{5/3}\frac{1}{(k_e^2+k^2\Lambda^2)^{17/6}}\exp\left(-2\frac{k^2}{k_d^2}\right) \label{eq:modifiedKarmanP}
\end{align}

The direct \Gaussian Transform is again given by Eq.~\eqref{eq:dirGaussTrans} and we write
\begin{multline}\label{eq:002}
\G(p(\sigma)) = \frac{1}{\sigma^2} \sqrt{\frac{\pi}{2\sigma^2}}\Bigg(\L^{-1}\Big(\frac{110}{27\pi}\frac{\Lambda^5 k_e^{5/3}}{(k_e^2+s\Lambda^2)^{17/6}}\\
\cdot \exp\left(\frac{-2s}{k_d^2}\right)\Big)(t)\Bigg)_{t=\frac{1}{2\sigma^2}}
\end{multline}
 This can be solved analytically, with the partial fraction expansion \cite[p.783]{bronstein_taschenbuch_2008}.
The image function is defined by $F(p) = H(p)/J(p)$, with $J(p)$ being a polynome about $p$. First, we determine the inverse \textsc{Laplace} function to $H(p)$ and $1/J(p)$. We get the inverse \textsc{Laplace} function by applying the convolution theorem afterwards.

The inverse Laplace function of $1/J(p)$ is given by Eq.~\eqref{eq:KarmanInvLaplace} as
\begin{align}
\L^{-1}\left(1/J(p)\right) &= \L^{-1} \left(\frac{1}{(p-\alpha)^{17/6}}\right)(t) \\
				&= \frac{e^{\alpha t} t^{11/6}}{\Gamma(17/6)}
\end{align}
and the inverse Laplace function of $H(p)$ yields
\begin{align}
\L^{-1}(H(p)) &= \L^{-1}_p \left(\exp(-2 \frac{p}{c^2})\right)(t) \\
			&=
\begin{cases} \delta\left(t-\frac{2}{c^2}\right)& \text{if } t \ge\frac{2}{c^2} \\
0 & \text{else.}
\end{cases}
\end{align}

So using the convolution theorem
\begin{align}
\L^{-1}\left(H(p)/J(p)\right) &= \L^{-1}(1/J(p)) * \L^{-1}(H(p)) \\
&=\begin{cases}\frac{1}{\Gamma(17/6)}(t-\frac{2}{c^2})^{11/6}\exp[\alpha (t-\frac{2}{c^2})]& \text{if } t \ge\frac{2}{c^2} \\
0 & \text{else.}
\end{cases} \notag
\end{align} 
Equation~\eqref{eq:002} simplifies to
\begin{align}
\G(p(\sigma))= \left\{
\begin{aligned}
\begin{split}
\frac{110\Lambda^{-2/3} k_e^{5/3}}{27\sqrt{2\pi}\Gamma(17/6)} \frac{1}{\sigma^3}
\left(\frac{1}{2\sigma^2}\!-\!\frac{2}{k_d^2}\right)^{\frac{11}{6}}\\
\cdot \exp{\left[-\frac{k_e^2 }{\Lambda^2}\left(\frac{1}{2\sigma^2}\!-\!\frac{2}{k_d^2}\right)\right]}
\end{split}
   &&\text{if } \frac{1}{2\sigma^2} \ge\frac{2}{k_d^2} \\
&0  &&\text{else.}
\end{aligned}
\right.\raisetag{15pt}
\end{align}
and we find the weighting function for the \modified \von \Karman spectrum as
\begin{align}
\boxed{
f_M(l) = \left\{
\begin{aligned}
\begin{split}
\frac{55\pi}{18\Gamma(17/6)}\sqrt[3]{\frac{k_e^{5}}{\Lambda^{2}}}\frac{1}{l^4}
\left(\frac{l^2}{\pi}-\frac{2}{k_d^2}\right)^{\frac{11}{6}}\\
\cdot \exp\left[-\frac{k_e^2 }{\Lambda^2}\left(\frac{l^2}{\pi}-\frac{2}{k_d^2}\right)\right]
\end{split}
	&& \text{if } l\ge\frac{\sqrt{2\pi}}{k_d} \\
&0  && \text{else.}
\end{aligned}\right.}\label{eq:weightmodKarman}
\end{align}
This function is plotted in Fig.~\ref{fig:weight}.
From the plot we see that the \modified \von \Karman and the \von \Karman weighting functions are identical in a region of length scales above the \textsc{Kolmogorov} scale. From the \modified \von \Karman weighting function a cut-off condition for the smalest relevant length scales can be derived as
\begin{align}
k_d\ge\frac{\sqrt{2\pi}}{l}.
\end{align}

\subsection{Weighting of velocity spectra}
Following Ref.~\cite{Pope_Turbulent_2000}, in isotropic turbulence the one-dimensional velocity spectra $E_{ii}(k_1)$ are determined by the energy-spectrum function $E(k)$:
\begin{align} \label{eq:VelSpec}
E_{ii}(k_1) = \iint\limits_{-\infty}^\infty \frac{E(k)}{2\pi k^2}\left(1-\frac{k_i^2}{k^2}\right) \abl k_2 \abl k_3.
\end{align}
As these integrals are independent of the integral length scale and the weighting function $f(l)$ is independent of the wavenumbers it is obvious that the realisation of arbitrary isotropic velocity spectra by superposition of \Gaussian velocity spectra is straight forward given by combining Eq.~\eqref{eq:weighting} and \eqref{eq:VelSpec} to
\begin{align}
E_{ii}(k_1,\Lambda) = \int\limits_0^\infty f(l,\Lambda) E_{ii}^G(k_1,l) \abl l.
\end{align}

\subsection{Two-dimensional turbulence}
The derivation has been performed in three-dimensional space. The application to two-dimensional turbulence is analogous. As the axial velocity spectrum $E_{11}(k_1)$ is identical in 2D and 3D space, the weighting functions $f(l)$ apply without modifications to 2D turbulence, resulting in a different one-dimensional transverse velocity spectrum $E_{22}(k_1)$ and energy-spectrum function $E(k)$ for 2D turbulence. 
%

\section{Discretisation}
\begin{figure}[h]
\centering
	\subfloat[Normalised energy \von \Karman spectrum realised by superposition of 10 weighted \Gaussian spectra compared to the analytical solution.\label{fig:resolution}]{\includegraphics[width=0.5\textwidth]{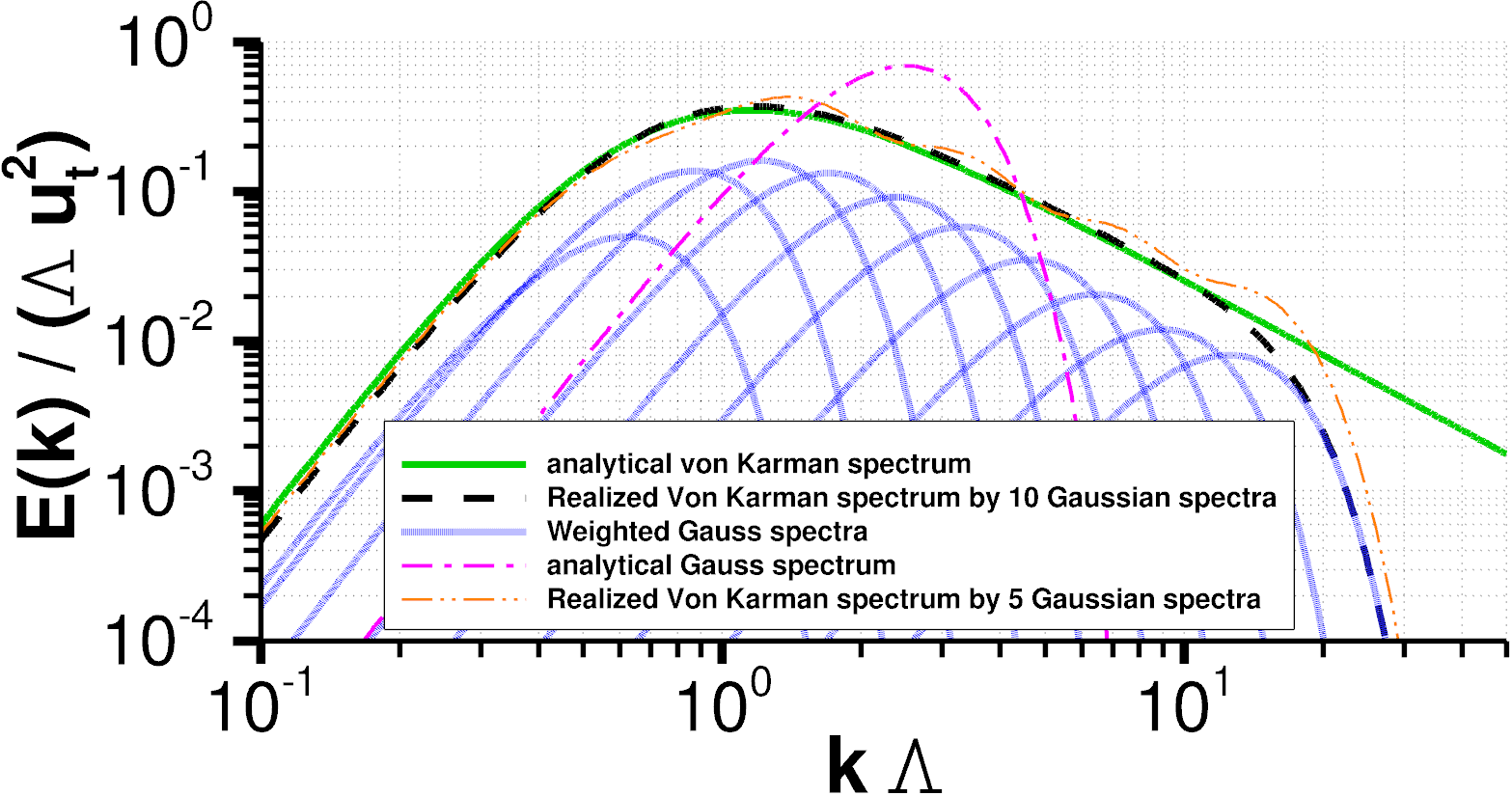}}\\
 	\subfloat[Power Spectral Density (PSD) of the axial turbulence velocity fluctuations realised by the RPM method in comparison to the measurements provided by Coupland \cite{Envia_AA-39_2014} and the analytical solution for a \von \Karman velocity spectrum.\label{fig:spectrum_uu}]{\includegraphics[width=0.5\textwidth]{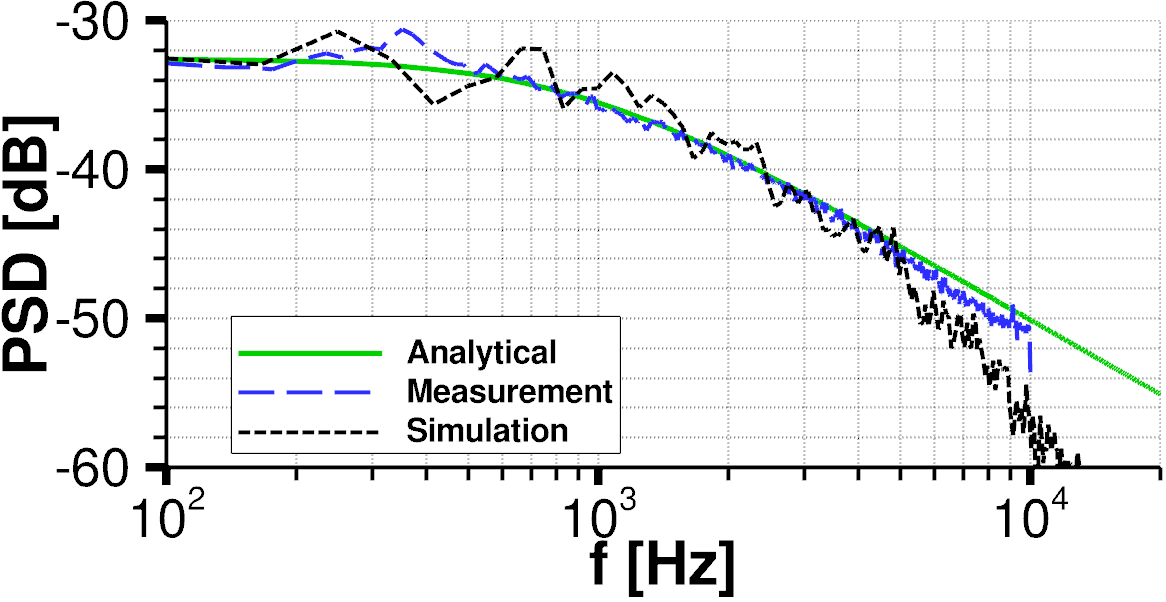}}
	\caption{The discretisation and application of the method to model turbulence spectra.}
\end{figure}
The main idea of the weighting function is to realise an arbitrary spectrum $E(k)$ of length scale $\Lambda$ with the superposition of a limited number of \Gaussian spectra $E_G$ of length scales $l_m$ with $0 \leq m\leq M$. For this we discretise the integral in Eq.~\eqref{eq:weighting} to
\begin{align}
E(k) &= \int\limits_0^\infty f(l) E_G(k,l) \abl l \\
     &\approx \sum\limits_{m=0}^{M} f(l_m) E_G(k,l_m) \Delta l_m \label{eq:discrete}
\end{align}
with the spacing $\Delta l_m$. 

For an efficient realisation we want $M$ to be as small as possible. As many orders of the wavelength $k$ have to be covered an exponential distribution of the length scales $l_m$ seems natural. We propose to discretise $l_m$ by 
\begin{align}
l_m = l_0 q^m, &&
q = \left(\frac{l_M}{l_0}\right)^\frac{1}{M}, && 
m = 0 \dots M 
\end{align}
 with the minimum and maximum length scales $l_0$ and $l_M$, respectively.
 To define the spacing $\Delta l_m$, the trapezoidal rule is applied:
\begin{align}
(\Delta l)_0 = \frac{l_{1}-l_{0}}{2}, &&
(\Delta l)_p = \frac{l_{p+1}-l_{p-1}}{2},&&
(\Delta l)_M = \frac{l_{M}-l_{M-1}}{2}.
\end{align}

 Analytical parameter variations show that a realisation with five filters per order of $k$-variation is already sufficient for to get a spectrum that is visually smooth. An example result is shown in Fig.~\ref{fig:resolution}. The green curve indicates the analytical \von \Karman spectrum and the black curve is the realisation with 10 weighted \Gaussian spectra for two orders of $k$-variation. The set of blue curves show the $M=10$ \Gaussian spectra of length scales $l_m$ weighted with the analytical weighting function $f_K(l_m)$ of Eq.~\eqref{eq:weightKarman}. In comparison a realisation with only 5 weighted \Gaussian spectra is plotted with the orange curve. The pink curve shows the original \Gaussian spectrum of Eq.~\eqref{eq:GaussEnergy} of the integral length scale $\Lambda$. 

Note that the smallest used length scale $l_1$ is chosen to resolve the highest wavenumbers of interest, but the length scale does not have to be smaller than the \textsc{Kolmogorov} length scale $l_d = \frac{2 \pi}{k_d}$.
For low wavenumbers no additional Gaussian spectra are needed as this region is efficiently covered if the largest length scale verifies $l_M \geq 4 \Lambda$. This is due to the fact that all investigated model spectra, including the \Gaussian spectrum, follow the power law $E(k) \sim k^4$ in the low wavenumber region.

As long as the 1D model spectra can be assumed valid, the findings are problem independent. Nevertheless depending on the wanted accuracy another discretisation or distribution density might be more appropriate and a quantification of the error is needed.

\section{Realisation in RPM}
Now we implement the analytical weighting functions in the RPM method of Ewert et al.~\cite{Ewert_CAA_2011}. The RPM method delivers unsteady fluctuating streamfunctions $ \pmb{\psi}$ and the turbulent velocity fluctuations $\mathbf v$ by convolution of mutually uncorrelated spatio-temporal white-noise $\pmb{\mathscr{U}}$ in a Lagrangian frame~\cite{Ewert_Broadband_2008} with the \Gaussian filter kernel G(r)

\begin{align}\label{eq:filter3D}
\mathbf v = \nabla \times \bm \psi =  \nabla \times\int \hat A\; G(\mathbf x - \mathbf x\:') \pmb{\mathscr{U}}(\mathbf x,t) \abl^d x',
\end{align}
realising a \Gaussian velocity spectrum.  The dimension of the problem is indicated by $d$ (2 for two-dimensional turbulence and 3 for three-dimensional turbulence). The variance of the fluctuations is inferred by Ref.~\cite{Ewert_CAA_2011} as
\begin{align}\label{eq:Eq34}
\hat R = \overline{\psi_i(\mathbf x, t)\psi_i(\mathbf x,t)} = \frac{\hat A(\mathbf x)^2 \Lambda^d}{\rho_0},
\end{align}
where $\rho_0$ is the local mean flow density. In order to realise a target kinetic energy $k_t = \frac{d}{2} u_t^2$ with length scale $\Lambda$ the source variance is defined as
\begin{align}
\hat R = \frac{ 2^{4-d}\Lambda^2 k_t}{3\pi}, \label{eq:variance}
\end{align} 
From Eq.~\eqref{eq:Eq34} the amplitude infers as
\begin{align}
\hat A = \sqrt{\frac{\rho_0 \hat R}{\Lambda^d}}.\label{eq:amplitude}
\end{align}

Now, to realise the discretised model spectra from Equations~\eqref{eq:weightKarman}, \eqref{eq:weightLiepmann} or \eqref{eq:weightmodKarman} the \Gaussian energy spectra $E_G(k,l_m)$ are weighted with $f(l_m) \cdot \Delta l_m$ and superposed as seen in Eq.~\eqref{eq:discrete}. Hence using Eq.~\eqref{eq:discrete},~\eqref{eq:amplitude} and ~\eqref{eq:variance} the amplitude for each discrete weighted \Gaussian energy spectrum $E_G(k,l_m)$ must be 
\begin{align}
\hat A_m = \sqrt{\frac{\rho_0}{l_m^{d-2}} \frac{ 2^{4-d} k_t}{3\pi}}.
\end{align}

Note that the correlations of the velocity fluctuations given in Equation~\eqref{eq:filter3D} perfectly match the complete correlation tensor of isotropic turbulence given in \cite[Equation~(9)]{Ewert_Broadband_2008}. Therefore the model spectra derived in this paper also realise isotropic solenoidal turbulence as they are a superposition of \Gaussian spectra.

\section{Application}
The analytical weighting has been successfully applied by the authors to reproduce the measured inflow turbulence spectrum used for the fundamental test case 1 of the AIAA benchmark workshop~\cite{Envia_AA-39_2014} with the RPM method. The energy spectrum discretisation shown in Fig.~\ref{fig:resolution} is the one used for the benchmark. Turbulent fluctuations of an integral length scale of $\Lambda = 8$mm  had to be resolved in the frequency range $100 Hz \leq f \leq 10kHz$. 
For a resolved \von \Karman spectrum in that range we needed $M = 10$ \Gaussian spectra reaching from length scales $l_0 = \Lambda/5$ to $l_M = 4\Lambda$. In Fig.~\ref{fig:spectrum_uu} the resulting measured and synthesised turbulent spectra of the axial velocity component are compared; the results show close agreement with the target spectrum. 

As each filtering process, generating the \Gaussian spectra of length scales $l_m$, is fully decoupled from the others, the method can be implemented thread-parallel. In this way the over-head for realising arbitrary spectra is easily handled by using additional computational power resulting in negligible penalty in efficiency compared to a computation realising a \Gaussian spectrum.

\section{Conlusion}
Analytical weighting functions have been derived to realise \von \textsc{K\'arm\'an}, \textsc{Liepmann}, and \modified \von \Karman spectra of integral length scale $\Lambda$ by integrating weighted \Gaussian spectra of different length scales. 

A discretisation of this integral is necessary in real applications.
Using an exponential distribution it was shown that very few \Gaussian realisations per order of frequency are enough for a smooth resulting spectrum. The discretisation is limited in the lower frequency range by $l\leq4\Lambda$ and the upper limit is either given by grid resolution or in case of the \modified \von \Karman spectrum by the Kolmogorov length scale $l_d=2\pi/k_d$.
 The method has been validated by generating a \von \Karman spectrum with the RPM method by a superposition of 10 \Gaussian realisations to resolve 2 orders of magnitude of the frequency range. 

\section{Acknowledgement}
The authors wish to acknowledge the financial support of the European Commission, provided in the framework of the FP7 Collaborative Project IDEALVENT (Grant Agreement no 314066).

\bibliographystyle{elsarticle-num}
\bibliography{bibliography}

\end{document}